\documentclass[twocolumn,prl]{revtex4}
\usepackage{epsfig,amssymb,amsmath}

\begin{document}
\title{Devil's Staircases and Continued Fractions in the Josephson Junctions}

\author{Yu. M. Shukrinov$^{1}$}
\author{S. Yu. Medvedeva$^{1,2}$}
\author{A. E. Botha$^{3}$}
\author{M. R. Kolahchi$^{4}$}
\author{A. Irie$^{5}$}
\address{$^{1}$ BLTP, JINR, Dubna, Moscow Region, 141980, Russia\\
$^{2}$Moscow Institute of Physics and Technology
 (State University), Dolgoprudny, Moscow Region, 141700, Russia\\
$^{3}$Department of Physics, University of South Africa, P.O. Box 392, Pretoria 0003, South Africa\\
$^{4}$Institute for Advanced Studies in Basic Sciences, P.O. Box 45195-1159, Zanjan, Iran\\
$^{5}$Department of Electrical and Electronic Systems Engineering, Utsunomiya University, 7-1-2 Yoto,
Utsunomiya 321-8585, Japan}

\date{\today}

\begin{abstract}
The detailed numerical simulations of the IV-characteristics of Josephson junction under external electromagnetic radiation show devil's staircases within different bias current intervals. We have found that the observed steps form very precisely continued fractions. Increasing of the amplitude of radiation shifts the devil's staircases to higher Shapiro steps. The algorithm of appearing and detection of the subharmonics with increasing radiation amplitude is proposed. We demonstrate that subharmonic steps registered in the famous experiments by A. H. Dayem and J. J. Wiegand [Phys. Rev 155, 419 (1967)] and J. Clarke [Phys. Rev. B 4, 2963 (1971)]  also form continued fractions.
\end{abstract}
\maketitle
Josephson junctions are regarded as excellent model systems for studying a variety of nonlinear phenomena in different fields of science \cite{barone82,likharev86} such as frequency locking, chaos, charge density waves, transport in superconducting nanowires, interference phenomena and others \cite{brown84,tekic11,hamilton72,kautz85}. These phenomena, and especially properties of the Shapiro steps (SS) \cite{shapiro63} in Josephson junctions  are very important for  technical applications \cite{kleiner04}.

In a Josephson system driven by an external microwave radiation, the so-called devil's staircase (DS) structure has been predicted as a consequence of the interplay of the Josephson plasma frequency, and the applied frequency (see Refs.\cite{benjacob81,jensen83} and references therein). To stress the universality in the scenario presented, we note that the devil's staircase appears in other systems including the infinite spin chains with long-range interactions \cite{nebendahl13}, frustrated quasi-two-dimensional spin-dimer system in magnetic fields \cite{tarigawa13},  systems of strongly interacting
Rydberg atoms \cite{weimer10}, and fractional quantum Hall effect \cite{laughlin85}. A series of fractional synchronization regimes (devil's staircase) in a spin-torque nano-oscillator driven by a microwave field was experimentally demonstrated \cite{urazhdin10}.  The devil's staircase is considered as an outstanding example of a `phase diagram' in physics, because it shows a high degree of self-organization \cite{yakes04}.

A detailed experimental investigation of the subharmonic SS  in SNS junctions were made by J.Clarke \cite{clarke71}. He found that the application to a junction of rf electromagnetic radiation of frequency $\Omega$ induced constant-voltage current steps at voltages $(n/m)\hbar \Omega/(2e)$, where $n$ and $m$ are positive integers. The results were explained based on the idea that phase difference in Josephson junction is increasing in time in a uniform manner and current-phase relation is nonsinusoidal. The junction generates harmonics when it biased at some voltage and these harmonics may synchronize with the applied radiation to produce the steps. Another famous experiment on the behavior of thin-film superconducting bridges in a microwave field by A. H. Dayem and J. J. Wiegand \cite{dayem67} also demonstrates the production of constant-voltage steps in the IV-characteristics. Some experimental results  are explained by nonsinusoidal current-phase relation \cite{bae08,kornev06}.  Ben-Jacob with coauthors \cite{benjacob81}  found the subharmonic steps within the resistively and capacitively shunted junction model (RCSJ) with purely sinusoidal current-phase relation \cite{stewart68,mccumber68}.

In this Letter we clearly show by high precision numerical simulations that IV-characteristic of a Josephson junction under microwave radiation exhibit DS structure of subharmonic Shapiro steps. To prove that we have a devil's staircase, we show its self-similar structure. The proof comes by analyzing the results in terms of the continued fractions \cite{khinchin64,cuyt08}.  We show that the steps observed in many previous experiments \cite{dayem67,clarke71,tarasov98,kautz94,kuznik93,seidel91,constantinian10,brown84} and numerical simulations \cite{benjacob81,tekic11,mali12,jensen83} form the continued fractions. We analyze the data of famous experiments of Clarke (see Ref.\cite{clarke71} and Fig. 9(a)) and  Dayem-Wiegand (see Ref.\cite{dayem67} and Fig. 16) in terms of continued fractions and show that the steps observed in these papers also form very precisely continued fractions.

Assuming the RCSJ model, we employ the following system of equations for the phase difference $\varphi$ across the junction, taking into account the external radiation with frequency $\omega$ and amplitude $A$:
\begin{eqnarray}
\dot{V}+\sin(\varphi)+\beta\dot{\varphi}=I+A\sin{\omega t}, \\
\dot{\varphi}=V.
\label{current}
\end{eqnarray}
Here the dc bias current $I$ and ac amplitude $A$ are normalized to the critical current $I_c$,  the voltage $V$ to the $V_0=\hbar \omega_p/(2e)$ ($\omega_p$ is the plasma frequency), time $t$ to the $\omega_p^{-1}$.  $\beta$ is the dissipation parameter ($\beta=\beta_c^{-1/2}$, $\beta_c $ is McCumber's parameter). In this study, we set $\beta=0.2$. Overdot indicates derivative with respect to the dimensionless time. In our simulations we used mostly 0.05 as a step in time, $10^{4}$ as a time domain for averaging with $10^3$ units before averaging,  $10^{-5}$ as a step in bias current. The details of simulation procedure are described in Ref.~\cite{sg-prb11}.
\begin{figure}[h!]
\includegraphics[height=33mm]{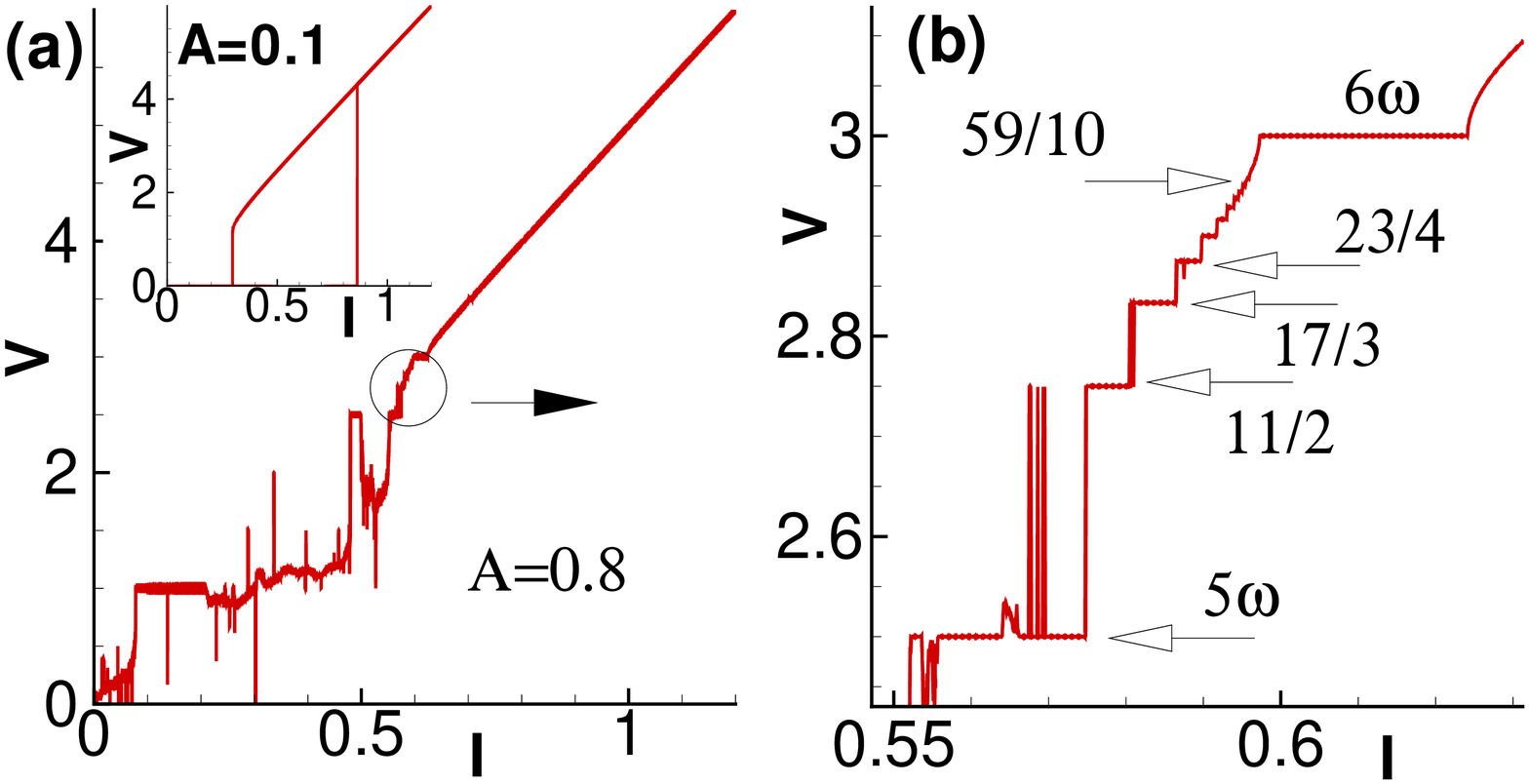}
\includegraphics[height=33mm]{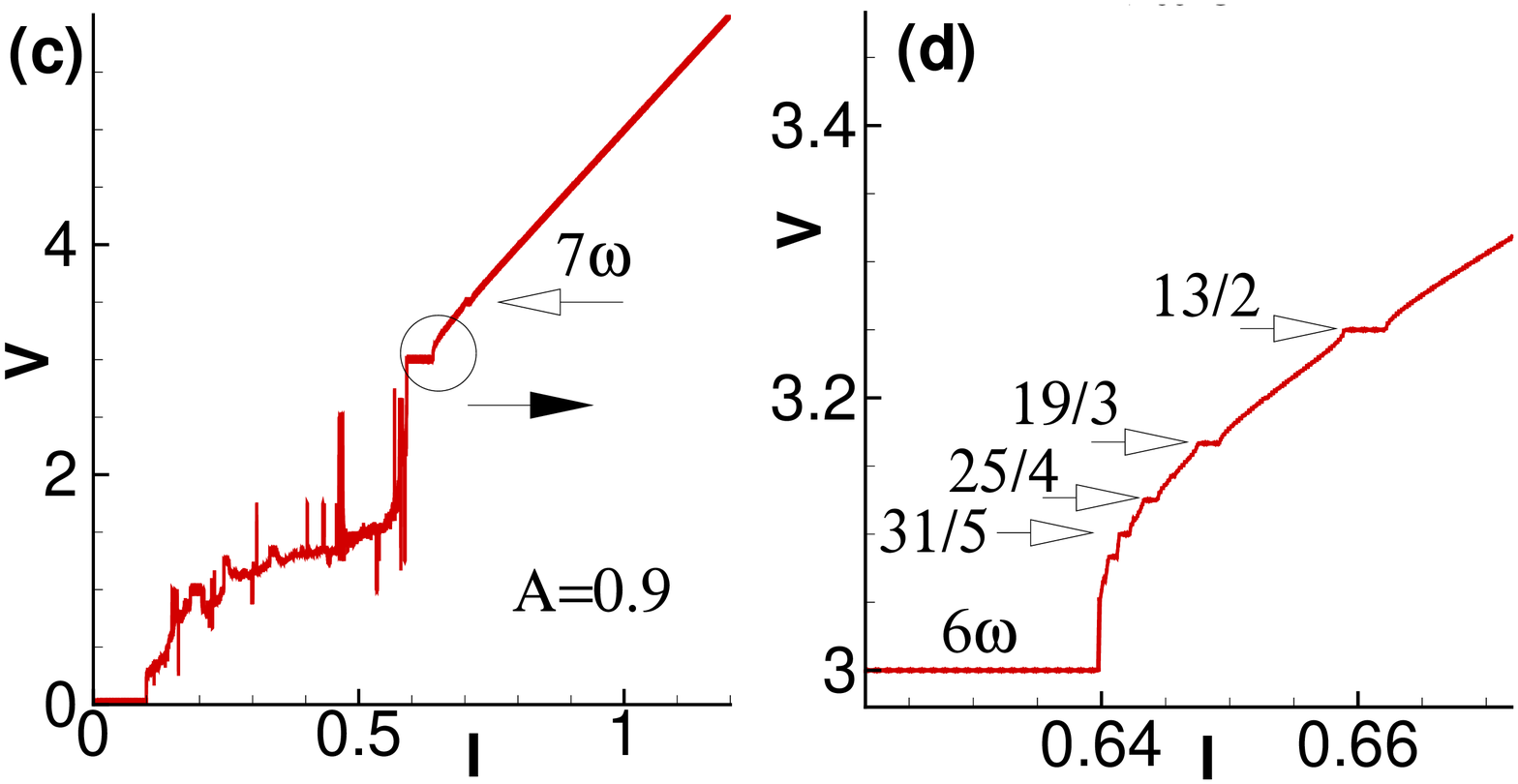}
\caption[bch!]{(Color online)  Simulated current-voltage characteristics of a Josephson junction under external electromagnetic radiation with $\omega =0.5$ and different radiation amplitudes $A$.  Figs. (b) and (d) show enlarged views of the encircled devil's staircases in Figs. (a), above, and (b), below, the 6th principle SS harmonic.}
\label{1}
\end{figure}

Figure ~\ref{1}(a) shows IV-characteristic of the Josephson junction at $\omega=0.5$ and $A=0.8$. We see that there is no hysteresis in comparison with the case at $A=0.1$ shown in the inset and chaos is developed in some current intervals. There is a manifestation of  the second harmonic, i.e. integer, Shapiro step at $V=2\omega=1$,  and the fifth and sixth, at $V=2.5$ and $V=3$, respectively. Let us consider carefully the part of IVC marked by circle which is  enlarged in Fig.~\ref{1}(b). A series of steps in the form of  $(N-1/n)\omega$, where $N=6$ and $n$ a positive integer, is observed  between $5\omega$ and $6\omega$.  We note that these  steps are approaching the 6-th harmonic from below. As $A$ is increased the chaos region is expanded and DS structure disappears. But instead it develops above the 6-th SS harmonic. Figure ~\ref{1}(c)  shows  IV-characteristic of the same Josephson junction at  $A=0.9$ with the DS structure  which is enlarged in Fig.~\ref{1}(d). The steps are approaching the  $6\omega$  harmonic from above and follow formula $(N+1/n)\omega$, again with $N=6$ and $n$ a positive integer.

The analysis of the various observed staircase structures leads us to the conclusion that in general the steps follow the formula for continued fractions, given by
\begin{eqnarray}
V=\left(N\pm\frac{1}{n\pm\frac{1}{m\pm\frac{1}{p\pm \ddots}}}\right)\omega,
 \label{cf}
\end{eqnarray}
where $N,n,m,p,\ldots $ are positive integers. We will call the terms that only differ in $N$, first-level terms. They describe Shapiro steps harmonics. The other terms describe the subharmonics, or the fractional steps.  Those differing in $N$ and $n$, we call  second-level terms; those differing in $N$, $n$ and $m$, third-level terms, etc.

Usually the mathematicians use the positive sign to express continued fractions \cite{khinchin64,cuyt08}. We have included the $'-'$ for convenience only; this allows us to easily analyze the subharmonics in the chosen interval of voltage (or frequency).  Another reason to use the continued fractions with negative sign is  following. The formula with positive signs puts the physically equal sequences
of subharmonics in different levels of formula. Consider the sequences 3/2,4/3,5/4... and 1/2,2/3,3/4... which describe the subharmonics placed on the same distance from the first Shapiro step; i.e. at $\omega$. In all plus continued fractions they are related to the different levels  described respectively by formulas  N+1/n and  $(N-1)+1/(n+1/m)$ with $N=1$ in first case and  $N=1, n=1$ in the second case. Including the $'-'$ allows us to use $N \pm 1/n$, with $'+'$ for the first and $'-'$ for the second sequence, and keeping $N=1$ for both sequences.
\begin{figure}[h!]
\includegraphics[height=50mm]{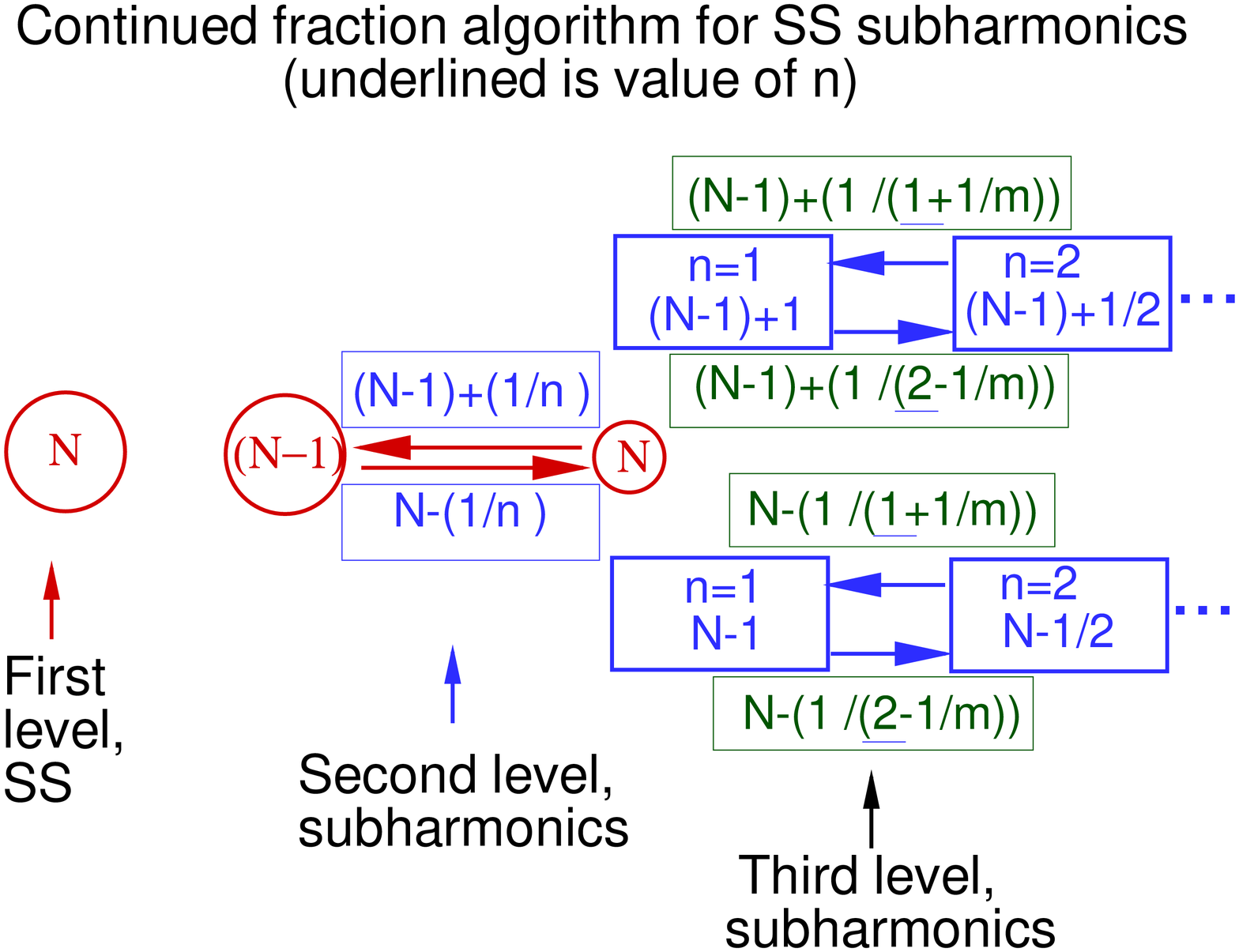}
\caption[bch!]{(Color online) Schematic demonstration of the appearance of continued fractions in IV-characteristic of Josephson junction under external electromagnetic radiation. $N$ is the SS number, $n$ and $m$ are positive integers.}
\label{2}
\end{figure}

The algorithm of continued fractions is schematically presented in Fig.~\ref{2}. We show by numbers in the circles the SS harmonics (red online). Second  level of continued fractions gives two groups of subharmonic steps (blue online): $(N-1)+(1/n)$ and $N-(1/n)$. First group is approaching $N-1$-th SS, and second one is approaching $N$-th SS.
So, if the sequence in the interval (a,b) is building to approach the step ``a'', need to take ``$+$'', and if the sequence is approaching the step ``b'',  then ``$-$''. To find subharmonics corresponding to the third level we first determine the interval  we are interested in; this entails, choosing $n$ and $n+1$, which are then kept constant, as $m$ is varied.  Each of them leads to the appearance of the other two groups, approaching the first and second term. In Fig.~\ref{2} we show the sequences of third level between the subharmonics with $n=1$ and $n=2$ also.  Other sequences are formed by the same algorithm.

\begin{figure}[h!]
\includegraphics[height=50mm]{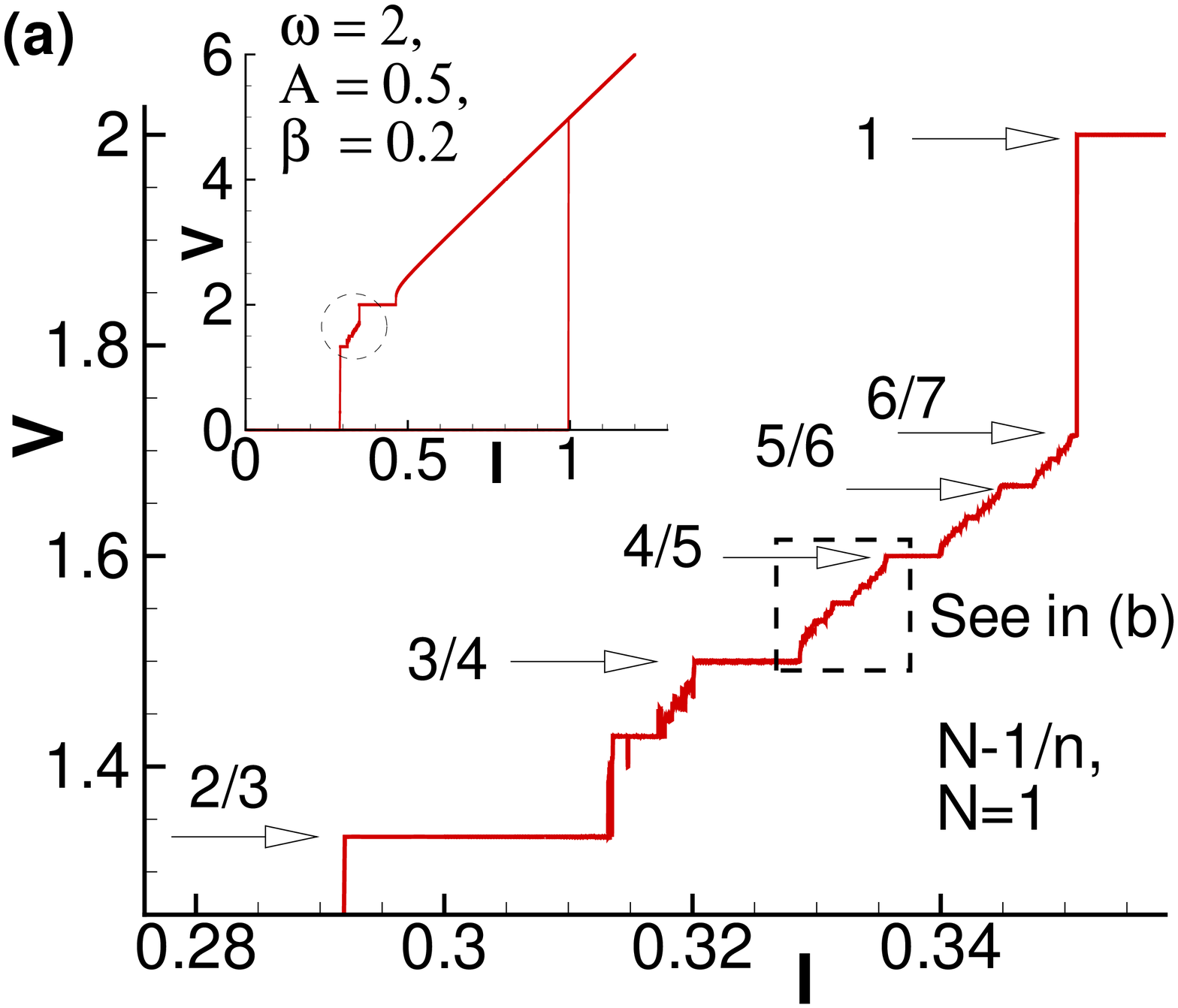}
\includegraphics[height=50mm]{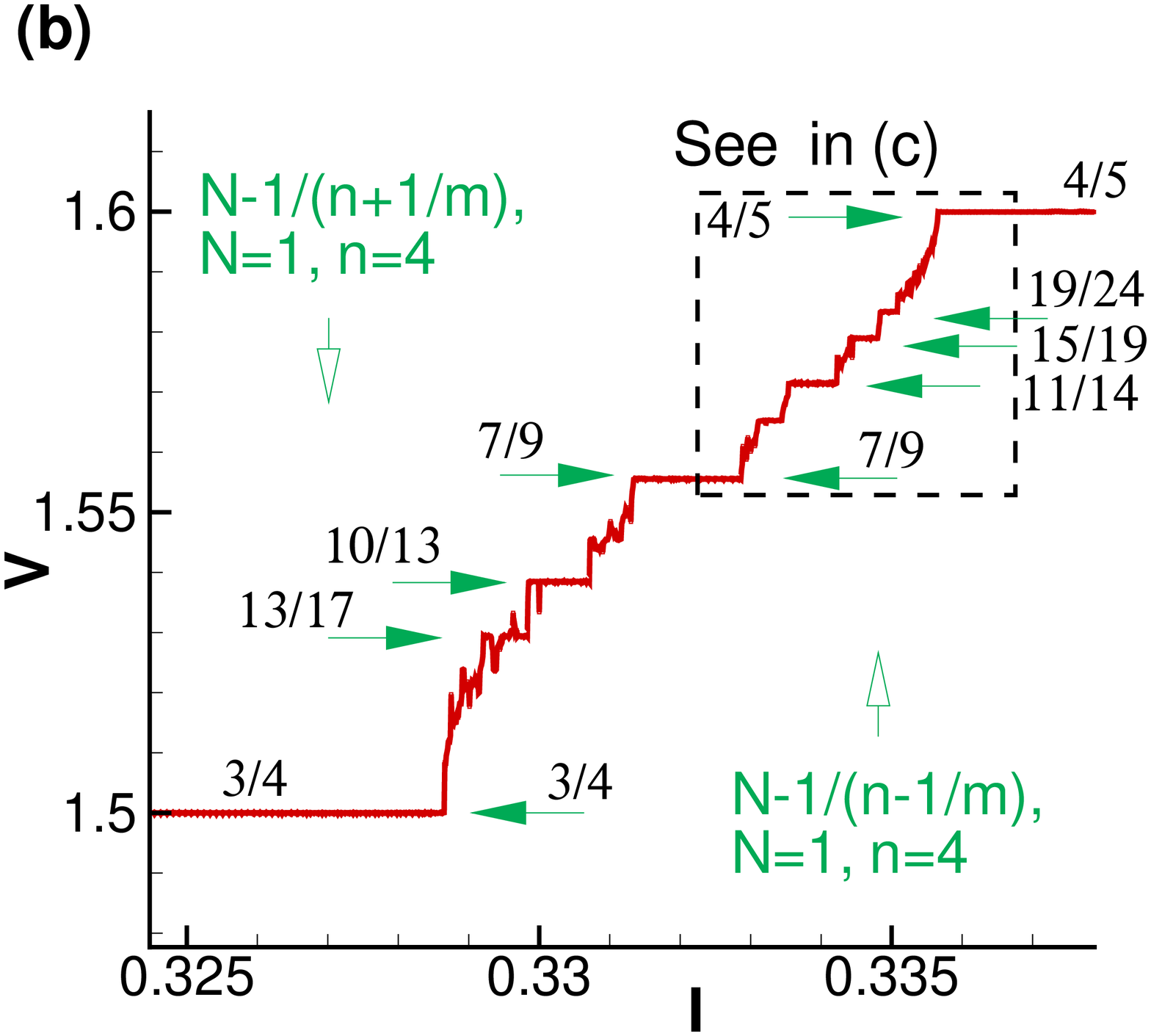}
\includegraphics[height=50mm]{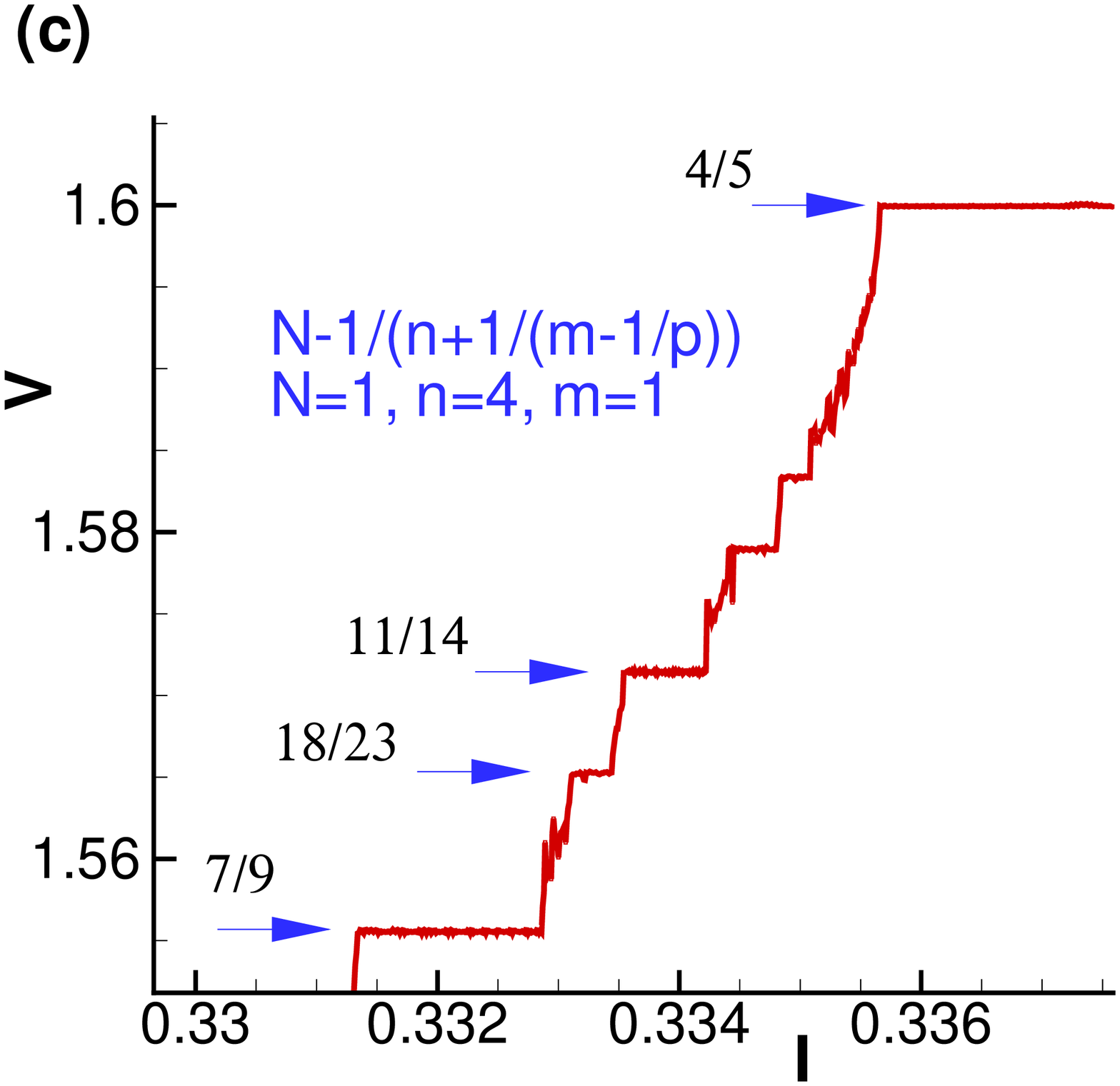}
\caption[bch!]{(Color online) The manifestation of the continued fractions in IV-characteristic of Josephson junction  at $\omega=2$ and $A=0.5$. (a) The steps in the interval between the zeroth and first SS; (b) The steps  between $\frac{3}{4}\omega$ and $\frac{4}{5}\omega$ marked by rectangle in (a); (c) The steps $\frac{7}{9}\omega$ and $\frac{4}{5}\omega$ marked by rectangle in (b). }
\label{3}
\end{figure}

Now, we set out to show the different levels of continued fractions of the devil's staircase. The DS in the IV-characteristic of the Josephson junction  at $\omega=2$ and $A=0.5$  is presented in  Fig.~\ref{3}. In one-loop IV-characteristic, shown in the inset to Fig.~\ref{3}(a), we see
that the return current is low enough to allow the $V=2$ step to develop.
The steps reflect the second level of the continued fractions  $(N-1/n)\omega$ with $N=1$. There is no half-integer step at $1/2$  in this IVC because of large value of the return current at chosen parameters.

The staircase bounded by the subharmonics, $3/4$ and $4/5$ and marked by a rectangle in Fig.~\ref{3}(a),  is enlarged in Fig.~\ref{3}(b). In particular, we see the sequence 4/5, 7/9, 10/13, 13/17..., reflecting the third level continued fraction $(N-1/(n+1/m))\omega$ with $N=1$, $n=4$ and the sequence 3/4, 7/9, 11/14, 15/19..., reflected   $(N-1/(n-1/m))\omega$ with $N=1$, $n=5$. Moreover,  the part between the steps $\frac{7}{9}\omega$ and $\frac{4}{5}\omega$ also marked by rectangle in this figure,  is enlarged in Fig.~\ref{3}(c). We found here the steps 7/9, 11/14, 15/19, 19/24, reflected the fourth level of continued fractions   $(N-1/(n+1/(m+1/p)))\omega$ with $N=1$, $n=4$, and $m=1$, and the sequence 4/5, 11/14, 18/23, reflected   $(N-1/(n+1/(m-1/p)))\omega$ with $N=1$, $n=4$, and $m=2$. Voltages found in our high precision numerical simulations, coincide with the corresponding values calculated by formula ~(\ref{cf}).

Let us finally discuss the experimental results on the subharmonic steps in IV-characteristic of a Josephson junction in presence of rf radiation. Our main statement is that the set of the constant voltage steps found in the previous experiments \cite{{clarke71},{dayem67},{tarasov98},{constantinian10}} are structured such that is reproduced by the continued fractions.

\begin{figure}[h!]
\includegraphics[height=50mm]{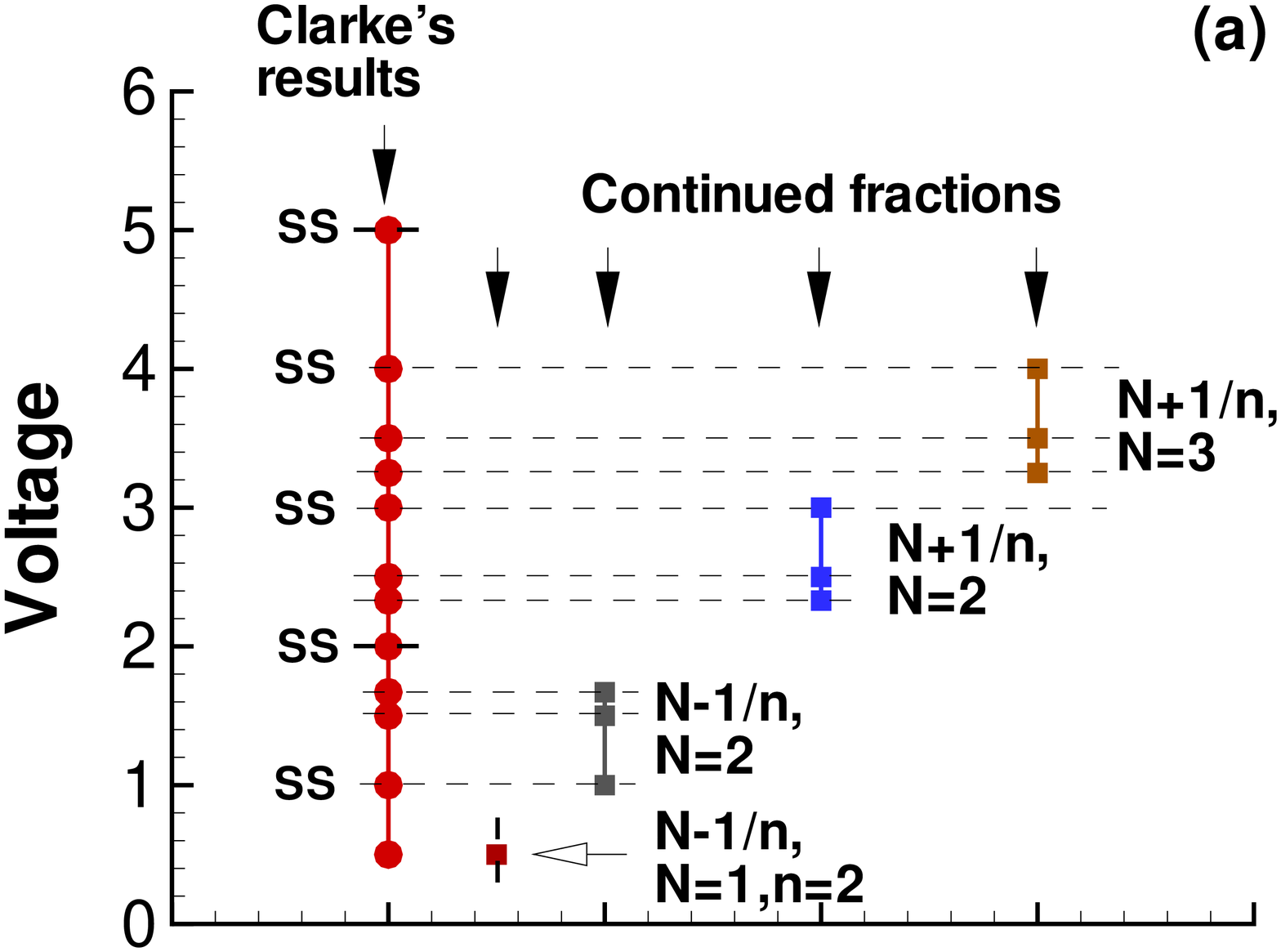}
\includegraphics[height=50mm]{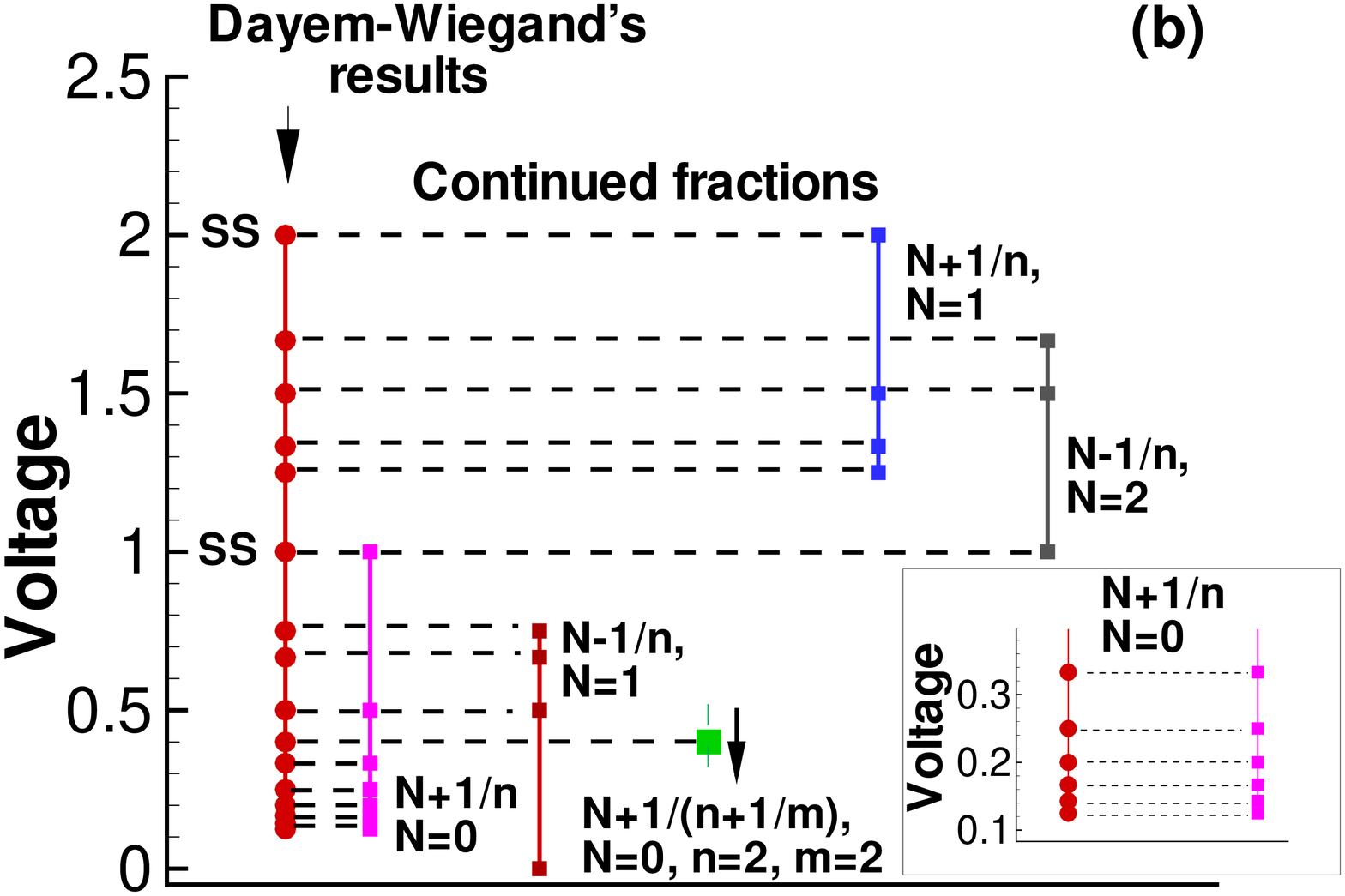}
\caption[bch!]{(Color online) Comparison of the experimental results of (a) Clarke\cite{clarke71} and  (b) Dayem and Wiegand\cite{dayem67}  with continued fractions. Filled circles show the experimental results, squares - different continued fractions.}
\label{4}
\end{figure}

We first consider the experiments of Clarke, and in particular look at Fig. 9(a) in Ref.\cite{clarke71}. In Fig.~\ref{4}(a) we reproduce these experimental results and compare them with continued fractions in the corresponding intervals of voltage. Voltage is normalized to the value of the first Shapiro step. In  the interval between the zeroth and first SS the subharmonic 1/2 is registered, reflecting the sequence $N+1/n$ with $N=0, n=2$. In the second SS interval (1,2) a series 1, 3/2, 5/3 is fixed which follow $V=(N-1/n)$ with $N=2$. In third (2,3) and forth (3,4) SS intervals  the steps at voltages 3/1, 5/2, 7/3 and 4/1, 7/2,..., 13/4 follow the fractions $V=(N+1/n)$ with $N=2$ and $N=3$, respectively. In the last series, it was only the 10/3 step that was not found.

The subharmonics which were  experimentally  measured by A. Dayem and J. Wiegand in Ref.\cite{dayem67} precisely follow the continued fraction formulas also. Figure 16 of  Ref.\cite{dayem67} shows the IV-characteristics at different power levels, for applied microwave radiation at $4.26$~GHz. In Fig.~\ref{4}(b) we also reproduce these experimental results and compare them with continued fractions.  The subharmonic steps in SS intervals (0,1) and (1,2) were found. The analysis shows that the   steps  0, 1/2, 2/3, 3/4 follow  $(N-1/n)$ with $N=1$  and the series 1/n is just  $(N+1/n)$ with $N=0$. For clarity we enlarge this part of figure in the inset.  In the SS interval (1,2) the experiment shows the steps 2/1, 3/2, 4/3, 5/4 according to $N+1/n$ with $N=1$, and 1, 3/2, 5/3 according to $N-1/n$ with $N=2$. It seems that there is a misprint in the original paper: the step around $V=4\mu V$ denoted as $1/5$. Actually, it is the step $2/5$ and it follows the third level of continued fractions $N+1/(n+1/m)$ with $N=0, n=2, m=2$. We see also in the analyzed  figure the signature of the step 3/5 between 1/2 and 2/3, followed $N-1/(n-1/m)$ with $N=1, n=3, m=2$, which was not marked by authors.

We note that in Ref.~\cite{tarasov98} the authors observed two
series of subharmonic steps up to sixth order ($n=6$)  experimentally. We consider these to be special cases of Eq.(~\ref{cf}): the first series corresponds to $V=(0+1/n)\omega $ and the second to $V=(1+1/n)\omega $.

Reports on measurements of dc electron transport and microwave dynamics of thin film hybrid Nb/Au/CaSrCuO/YBaCuO planar Josephson junctions were presented in Ref.~\cite{constantinian10}. The authors observed tunnel-like behavior, and oscillations in sync with the applied radiation at integer and half-integer steps. For a junction  fabricated on \emph{c-}oriented YBCO film the devil's staircase structure was observed under microwave irradiation at $4.26$~GHz.

In summary, the detailed numerical simulations of the IV-characteristic of a Josephson junction under microwave radiation allowed us to demonstrate  a self-similar structure of Shapiro steps subharmonics known as a devil's staircase.  We conclude that in many experimental and simulated physical systems, in various fields, the response function of the driven system has the devil's staircase structure, characterized by a continued fraction.

Yu. M. S. thanks  I. Rahmonov, M. Yu. Kupriyanov, K.Y. Constantinian,  G. A. Ovsyannikov  for helpful discussions and D. V. Kamanin and the JINR-SA agreement for the support of this work. He also appreciates kind hospitality of Prof. Y. Takayama and Prof. N. Suzuki from Utsunomiya university where part of this work was done.


\begin{thebibliography}{10}
\bibitem{barone82}Antonio Barone and Gianfranco Paterno, Physics and Applications of the Josephson Effect, John Wiley and Sons Inc., 1982.
\bibitem{likharev86} K. K. Likharev, Dynamics of Josephson Junctions and Circuits, Gordon and Breach, Philadelphia, 1986.
\bibitem{brown84}S. E. Brown, G. Mozurkewich and G. Gruner, Phys. Rev. Lett. {\bf 52}, 2277 (1984).
\bibitem{tekic11} J. Tekic, Z. Ivic, Phys. Rev. E {\bf 83}  056604 (2011).
\bibitem{kautz85} R. L. Kautz and R. Monaco, J. Appl. Phys. {\bf 57}, 875 (1985).
\bibitem{hamilton72}C. A. Hamilton and E. G. Johnson Jr., Physics Letters A {\bf 41}, 393  (1972).
\bibitem{shapiro63} S. Shapiro, Phys. Rev. Lett. {\bf 11}, 80 (1963).
\bibitem{kleiner04}W. Buckel and R. Kleiner, Superconductivity: Fundamentals and Applications (Wiley-VCH, Weinheim, 2004).
\bibitem{jensen83}M. H. Jensen, P. Bak and T. Bohr, Phys. Rev. Lett. {\bf 50}, 1637 (1983).
\bibitem{benjacob81} E. Ben-Jacob, Y. Braiman, R. Shainsky, Appl. Phys. Lett. {\bf 38}, 822 (1981).
\bibitem{nebendahl13} V. Nebendahl and W. Dür, Phys. Rev. B {\bf87}, 075413 (2013).
\bibitem{tarigawa13} M. Takigawa et. al., Phys. Rev. Lett. 110, 067210 (2013)
\bibitem{weimer10} H. Weimer and H. P. Büchler, Phys. Rev. Lett. 105, 230403 (2010)
\bibitem{laughlin85}R. B. Laughlin et al., Phys. Rev. B {\bf 32}, 1311 (1985).
\bibitem{urazhdin10}S. Urazhdin at al., Phys. Rev. Lett. 105, 104101 (2010)
\bibitem{yakes04}M. Yakes, V. Yeh, M. Hupalo, and M. C. Tringides, Phys. Rev. B {\bf69}, 224103 (2004).
\bibitem{clarke71}J. Clarke, Phys. Rev. B  {\bf 4}, 2963 (1971).
\bibitem{dayem67}A. H. Dayem, J. J. Wiegand, Phys. Rev.  {\bf 155}, 419 (1967).
\bibitem{bae08}Myung-Ho Bae et al., Phys. Rev. B {\bf77}, 144501 (2008).
\bibitem{kornev06} V. K. Kornev et al., Physica C {\bf435}, 27 (2006).
\bibitem{stewart68}C. Stewart, Appl. Phys. Lett. {\bf12}, 277 (1968).
\bibitem{mccumber68}D. E. McCumber, J. App\. Phys. {\bf39},3113 (1968).
\bibitem{khinchin64}A. Ya. Khinchin,  Continued Fractions. University of Chicago Press, 1964.
\bibitem{cuyt08}A. Cuyt, V. Brevik Petersen, B. Verdonk, H. Waadeland, W. B. Jones, Handbook of Continued fractions for Special functions, Springer Verlag, 2008.
\bibitem{tarasov98}M. Tarasov et. al., JETP Letters, {\bf68}, 454 (1998).
\bibitem{constantinian10}K. Y. Constantinian et al., JPCS, {\bf 234}, 042004 (2010).
\bibitem{kuznik93}J. Kuznik, K. Rogacki,  Physics Letters A {\bf 176}, 144 (1993).
\bibitem{seidel91}P. Seidel, M. Siegel and E. Heinz, Physica C {\bf 180}, 284 (1991).
\bibitem{kautz94} R. L. Kautz, S. B. Benz, C. D. Reintsema, Appl. Phys. Lett. {\bf 65}, 1445 (1994).
\bibitem{mali12}P. Mali et al., Phys. Rev. E {\bf 86},  046209 (2012).
\bibitem{sg-prb11}Yu. M. Shukrinov and M. A. Gaafar.  Phys. Rev. B {\bf 84}, 094514 (2011).
\end{thebibliography}
\end{document}